%% file: main.tex
\documentclass[12pt]{scrartcl}
\usepackage[top=2.5cm, bottom=2.5cm, left=2.1cm, right=2.1cm]{geometry}
\usepackage{amsmath,amssymb}  
\usepackage{enumerate}
\usepackage{graphicx} 
\usepackage{hyperref}
\hypersetup{colorlinks=false}
\usepackage{subfigure}
\usepackage{bm}

\usepackage[titletoc,page]{appendix}

\usepackage[american]{babel}
\usepackage[T1]{fontenc}
\usepackage{lmodern}
\usepackage[utf8]{inputenc}  

\usepackage{array}
\newcolumntype{C}[1]{>{\centering\arraybackslash}p{#1}}

\usepackage{dirtytalk}
\usepackage{etoolbox}

\usepackage[thmmarks,amsmath,hyperref,noconfig]{ntheorem}

\theorembodyfont{\upshape}

\theoremstyle{nonumberplain}
\theoremheaderfont{\itshape}
\theorembodyfont{\normalfont}
\theoremseparator{.}
\theoremsymbol{\ensuremath{_\blacksquare}}

\qedsymbol{\ensuremath{_\blacksquare}}

\renewrobustcmd{\bfseries}{\fontseries{b}\selectfont}

\usepackage{color,colortbl}
\definecolor{silke}{rgb}{0.0,0.0,1.0}
\definecolor{misha}{rgb}{1,0.6,0.0}
\definecolor{ariel}{rgb}{1.0, 0.44, 0.37}
\definecolor{david}{rgb}{0.0,0.7,0.3}
\definecolor{todo}{rgb}{1.0,0.0,0.0}

\definecolor{darkred}{rgb}{0.3,0,0}
\definecolor{darkgreen}{rgb}{0,0.3,0}
\definecolor{darkblue}{rgb}{0,0,0.3}
\definecolor{pink}{rgb}{0.78,0.09,0.51} 

\definecolor{orange}{rgb}{1,0.6,0.0}
\definecolor{grey}{rgb}{0.4,0.4,0.4}
\definecolor{lightgray}{gray}{0.9}
\definecolor{aquamarine}{rgb}{0.4,0.8,0.65}

\newcommand{\R}{\mathbb{R}}
\newcommand{\N}{\mathbb{N}}
\newcommand{\E}{\mathbb{E}}

\newcommand{\GP}{{\mathcal G \mathcal P}}
\newcommand{\var}{\text{Var}}
\newcommand{\cov}{\text{Cov}}
\newcommand{\corr}{\text{Corr}}
\newcommand{\U}{\bm{U}}
\newcommand{\su}{\bm{u}}

\newcommand{\bv}[1]{\boldsymbol{#1}}
\newcommand{\xb}{\bm{\mathrm{x}}}  % bold roman x (for vector)
\newcommand{\x}{\bm{\mathrm{x}}}  % bold roman x (for vector)

\usepackage{nomencl}
\makenomenclature

\usepackage{comment}

\newcommand{\citep}{\cite}
\newcommand{\upi}{\pi}

\begin{document}

\title{Global Stochastic Optimization of Stellarator Coil Configurations}
\date{}
% authors in alphabetical order
\author{\normalsize{\textbf{Silke Glas$^{1}$, Misha Padidar$^{2}$, Ariel Kellison$^{1}$ and David Bindel$^{1}$}}}
%\author{Author One Author Two}
\date{\small{$^{1}$Department of Computer Science, Cornell University, Ithaca, NY 14853, USA \\$^{2}$Center for Applied Mathematics, Cornell University, Ithaca, NY 14853, USA.}}

\maketitle 

%------------------------------------ABSTRACT ---------------------------------------------------------------------------
\begin{abstract}
\noindent
In the construction of a stellarator, the manufacturing and assembling of the coil system is a dominant cost. These coils need to satisfy strict engineering tolerances, and if those are not met the project could be canceled as in the case of the National Compact Stellarator Experiment (NCSX) project \citep{orbach2008statement}. 
Therefore, our goal is to find coil configurations that increase construction tolerances without compromising the performance of the magnetic field.

In this paper, we develop a gradient-based stochastic optimization model which seeks robust stellarator coil configurations in high dimensions.
In particular, we design a two-step method: first, we perform an approximate global search by a sample efficient trust-region Bayesian optimization; second, we refine the minima found in step one with a stochastic local optimizer.  To this end, we introduce two stochastic local optimizers: BFGS applied to the Sample Average Approximation and Adam, equipped with a control variate for variance reduction. 
Numerical experiments performed on a W7-X-like coil configuration demonstrate that our global optimization approach finds a variety of promising local solutions at less than $0.1\%$ of the cost of previous work, which considered solely local stochastic optimization.
\end{abstract}

\input{./sections/introduction.tex}
\input{./sections/focus.tex}
\input{./sections/problem_formulation.tex}
\input{./sections/global_optimization.tex}
\input{./sections/experiments.tex}
\input{./sections/conclusion.tex}

%%%%%% REFERENCES %%%%%% - to be included
\bibliographystyle{abbrv}
\bibliography{riskstell_draft}

\end{document}

%% file: sections/introduction.tex
\section{Introduction}\label{Sec:Introduction}

% In order to find good stellarator configurations, the design process is usually split into two steps.
The design process of finding promising stellarator coil configurations is traditionally split into two steps.
First, one aims to find the optimal plasma shape with respect to performance criteria such as, e.g., the magnetohydrodynamic (MHD) stability or alpha particle confinement. 
For the second step - the coil design - one tries to reproduce the target magnetic field confining the plasma.
Construction and placement of these coils are difficult tasks since minor errors in the fabrication or alignment can lead to major modifications of the magnetic field and thus poor particle confinement \citep{andreeva2009influence,lobsien2018stellarator}. 

\begin{figure}
    \centering
	\includegraphics[scale=0.3]{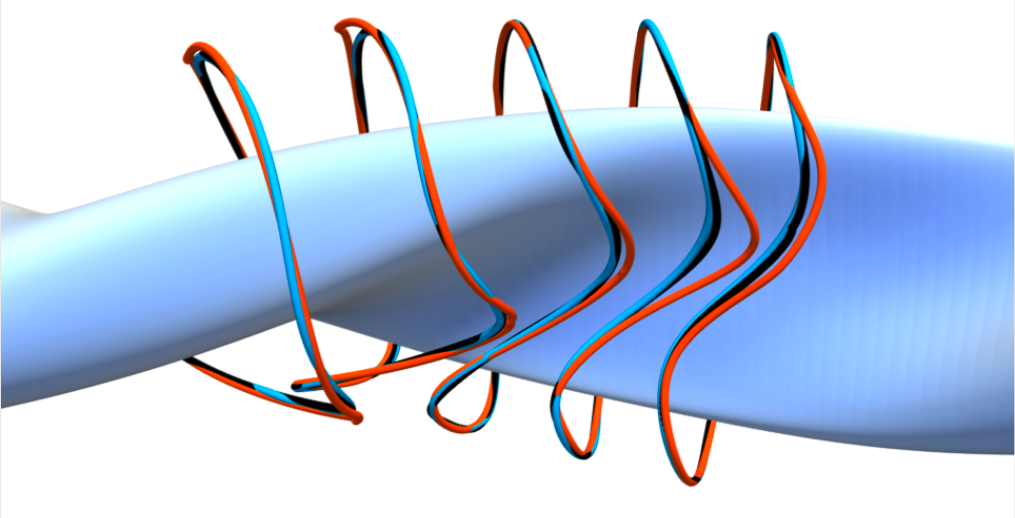}
	\caption{Variety of stochastic minima derived with DTuRBO and AdamCV.}
\label{Fig:W7-X_coils_comparison}
\end{figure}

To be able to guarantee good confinement when the stellarator is in operation, stringent engineering tolerances on the coils are required during the stellarator fabrication and assembly process. Unfortunately, such stringent tolerances can increase the cost and production timeline of a stellarator, 
which led to cancellation of the National Compact Stellarator Experiment (NCSX) project \citep{orbach2008statement}. During the Columbia Nonneutral Torus (CNT) design, a stochastic perturbation analysis was used to identify the robustness of coil-induced magnetic fields to coil alignment errors \citep{kremer2007creation}. This allowed the reduction of engineering coil tolerances significantly \citep{kremer2003status}. Furthermore, recent work has shown that stochastic optimization is a promising method to improve the robustness of the generated magnetic field to fabrication and alignment errors in the associated coils \citep{lobsien2020,wechsung2021singlestage}. However, all aforementioned approaches do not globally explore the coil design space. Indeed, design engineers are usually interested in having not only \emph{one}, but \emph{multiple} designs to choose from. In addition to the magnetic field, engineers care about other physical properties such as specific aspect ratio and rotational transform; see Figure \ref{Fig:W7-X_coils_comparison} for an example of different coils. Thus, there is a need for efficient stochastic methods for global exploration in order to provide multiple coil configurations in the design process. 
% Thus, we need more efficient stochastic methods for global exploration in order to provide multiple coil configurations in the design process. 

In this work, we develop an algorithm for the efficient \emph{global} stochastic optimization of stellarator coil configurations. We seek a coil set that is robust to random errors in the coils in expectation:
\begin{subequations} \label{eq:main_1}
\begin{align}
\min_{\x \in \R^{N}} \ \ &\E[f_\text{stoc}(\xb+\U)] + f_{\text{reg}}(\x) \\
% &c(\x) \ge \epsilon,
%\\
& c_i(\xb) \ge \epsilon_i \quad i=1,\ldots,N_C-1
\end{align}
\end{subequations}
where $\x\in\R^N$ defines the geometry of the coil set, $\U$ is a random error to the coils, $\E$ is the expectation over $\U$, $N_C$ is the number of coils in a single field period, $\epsilon \in \R$ is a constraint value, and the functions $f_\text{stoc}, f_{\text{reg}}, c_i:\R^{N} \rightarrow \R$ measure the quality of the coils and induced magnetic field. We define the components of the optimization model precisely in Section \ref{Sec:Problem_Formulation}. Similarly to \citep{lobsien2020}, we consider fabrication errors to be spatially correlated Gaussian perturbations to the coils.
In order to optimize \eqref{eq:main_1}, we present a two-step global-to-local algorithm for stochastic optimization. In the first step, we perform a global exploration of the stochastic optimization model given in \eqref{eq:main_1} using a trust-region Bayesian optimization method based off \citep{Turbo2019}. This global stage finds multiple approximate minima which, in the second step, are resolved by local stochastic optimizers. To perform local stochastic optimization we apply the BFGS optimizer \citep{nocedal2006numerical} to the Sample Average Approximation (SAA) \citep{shapiro2001monte} of \eqref{eq:main_1}  as well as the Adam optimizer \citep{kingma2014adam} enhanced by a control variate for variance reduction. Finally, we arrive at multiple local stochastic minima from which one can choose the most promising configuration. 

From a stellarator optimization point of view, the main new ingredients of our optimization routine are (to the best of our knowledge):
\begin{enumerate}
%	\item First \emph{stochastic optimization} in coil optimization code \emph{FOCUS} \citep{Zhu_2017}.
%	\item \emph{Coil-to-coil constraint} in FOCUS instead of previous coil-to-coil separation. 
	\item Method for \emph{efficient stochastic global exploration} of the design space. 
	\item Local stochastic optimization with Adam enhanced with novel \emph{control variate}.
\end{enumerate}
Moreover, this is the first work to perform stochastic optimization using the coil optimization code \emph{FOCUS} \citep{Zhu_2017}. 
%Additionally, we add a new \emph{coil-to-coil constraint} in FOCUS instead of the previous coil-to-coil separation. 

This paper is organized as follows. We briefly describe FOCUS and the internal representation of the coils in Section \ref{sec:focus}. In Section \ref{Sec:Problem_Formulation}, we introduce the formulation of the stochastic optimization model. Subsequently in Section \ref{Sec:global_high_dim_optimization}, we describe our global-to-local stochastic optimization algorithm. Numerical experiments on a W7-X configuration are presented in Section \ref{Sec:Experiments}. We conclude with a summary and ideas for future work in Section \ref{Sec:Conclusion}. 

%% file: sections/focus.tex
%!TEX root = ../main.tex

%------------------------------------SECTION: PROBLEM FORMULATION --------------------------------------
\section{FOCUS}\label{sec:focus}
%--------------------------------------------------------------------------------------------------------------------------------
%In this section we describe how FOCUS is used for coil optimization. We start with a description of how FOCUS represents the coils and subsequently explain how the objective function is depicted.   

%\subsection{Coil Representation}
For the second design stage of a stellarator, multiple codes are available to optimize for coil configurations which replicate a target magnetic field. Well-known examples are NESCOIL \citep{Merkel_1987}, REGCOIL \citep{Landreman_2017}, ONSET \citep{drevlak99}, COILOPT \citep{doi:10.13182/FST02-A206}, COILOPT++ \citep{7482426} and FOCUS \citep{Zhu_2017}.
In order to perform stochastic optimization, we decided to use FOCUS for two reasons: first, it allows coils to move freely in space whereas the other aforementioned coil optimization codes restrict to a so-called \emph{winding surface}; and second, FOCUS provides \emph{analytic first-order derivatives} which improve efficiency in the optimization process.%\footnote{Note, that although we make use of the analytic derivatives here, those are not necessary for the proposed algorithm.} 

In FOCUS, the coils are described by a 3-dimensional Fourier representation. FOCUS takes as input a vector of Fourier coefficients $\xb = (\x^{1}, \ldots, \x^{N_C} ) \in  \R^{N}$ describing the geometry of $N_C$ filamentous coils. From the Fourier coefficients $\xb^i = (x^i_{c,0},\ldots, x^{i}_{s,N_F}, y^i_{c,0},\ldots, y^{i}_{s,N_F}, z^i_{c,0}, \ldots,  z^{i}_{s,N_F})^{T}$ for the $i$-th coil, FOCUS constructs coils as a parametrized curve with $x$-coordinate
\begin{align*}
x^i(\x^{i},t) = x^i_{c,0} + \sum_{n=1}^{N_F} \left[ x^i_{c,n}\cos(nt) + x^i_{s,n}\sin(nt) \right], \qquad t \in [0,2\upi), 
%\label{eq:focus_fourier}
\end{align*}
with $N_F$ being the number of Fourier modes and analogous forms for the $y$-,$z$-coordinates $y^{i}(\x,t)$, $z^{i}(\x,t)$.\footnote{Note that the coil currents are held constant in this work.} By $\bm{X}^i(\xb,t) = (x^i(\x,t),y^i(\x,t),z^i(\x,t))^{T}$ we denote the parametric representation of the $i$-th coil filament with $t\in[0,2\upi)$. The number of parameters per coil is given by $3(2N_F +1)$ such that the total number of parameters results in $N:= 3N_C(2N_F +1).$ Note, that FOCUS only requires the description of a single field period of coils, constructing the other coils through reflection and rotation around the stellarator. 

With the Fourier representation FOCUS can, among other things, efficiently compute coil metrics related to coil curvature, coil length, normal magnetic field, quasisymmetry, magnetic island width and their respective gradients with respect to the Fourier coefficients \cite{zhu2019identification}. In this study, we only make use of the normal component of the magnetic field $f_{B} $ and length metric $f_{L}$, which we describe in the following.

The metric $f_{B}$ addresses deviations in the magnetic field produced by a coil set $\x$ compared to a target magnetic field $\bv{B}$.
Given a target plasma boundary $S$ and a target magnetic field $\bv{B}$ the normal field error $f_B$ is given by
\begin{equation}
    f_B(\bv{\xb}) = \int_S \frac{1}{2} (\bv{B}\cdot\bv{n})^2 ds,
    \label{Eq:average_squared_FE}
\end{equation}
where $\bv{n}$ is the unit normal on the target plasma boundary. Typically the target plasma boundary is generated from a MHD equilibrium solver such as e.g., VMEC \citep{hirshman1983steepest}, SPEC \citep{SPEC}.  
%These attributes make FOCUS a powerful tool for finding Stellarator coil-configurations that achieve a target plasma boundary.
%The dominant criterion $f_B$ in the objective $f$ addresses deviations in the magnetic field produced by a coil set $\xb$ from a target magnetic field. For a target normal magnetic field, the objective on the magnetic field $\bm{B}$ external to the plasma  generated by the coil set $\xb$ is defined as
%\begin{align}\label{Eq:average_squared_FE}
%    f_B(\bv{\x}) = \int_S \frac{1}{2} (\bm{B}\cdot\bm{n})^2 dS,
%\end{align}
%where $\bm{n}$ is the unit normal on the plasma boundary and $S$ is the area of the plasma boundary. 

%Inclusion of a penalty for coil length promotes coil sets that are desirable from an engineering perspective, namely coils that are not prohibitively long. 
The metric $f_L$ is introduced to influence the length of the coils such that they are not prohibitively long. 
For the $i$-th coil with length $L_i(\xb) \in \R$ and user-specified target length $L_{i}^{\text{target}}$, FOCUS computes the length metric
\begin{align}
    f_L(\bv{\xb}) = \frac{1}{N_C} \sum_{i=1}^{N_C}\frac{1}{2}\frac{(L_i(\x) - L_{i}^{\text{target}})^2}{\left(L_{i}^{\text{target}}\right)^2}.
    \label{eq:length_objective}
\end{align} 
In the next section, we detail how we design our stochastic optimization model using the introduced FOCUS metrics. 
 %In the following we detail how we perform stochastic optimization on this objective function. 

%% file: sections/problem_formulation.tex
\section{Optimization Model Formulation}\label{Sec:Problem_Formulation}

In this section we describe our stochastic optimization model for finding a set of coils which generate a target magnetic field and simultaneously hedge against errors in the coil fabrication. The decision variables $\xb\in \R^{N}$ for our problem are the $N = 3N_C(2N_F +1)$ Fourier coefficients defining the geometry of the $N_C$ coils, see Section \ref{sec:focus}. The stochasticity is motivated by fabrication errors in the coils and modeled as spatially correlated Gaussian perturbations as described in Section \ref{sec:uncertainty}. In Section \ref{sec:objectives} we detail the stochastic and non-stochastic parts of the objective function. We design a coil-to-coil separation constraint and formulate the final optimization problem used in the numerical experiments in Section \ref{sec:constraints}. 

%The weights used in our experiments as well as other optimization parameters for FOCUS can be found in Table \ref{Table:focus_opt_params}. Heuristically, we found that these weights struck a nice balance between minimizing field error and finding smooth coils. 

%Similar to \citep{lobsien2020, wechsung2021singlestage} 

\subsection{Coil Fabrication Uncertainty}
\label{sec:uncertainty}
% While the fabrication errors are typically thought to not effect the center position of the coil or orientation, they do alter the shape of the coil and in turn  the placement.
Errors during coil fabrication can alter the shape of the coil and in turn the relative positioning of the coil to the plasma. 
Mathematically, we model coil fabrication errors as spatially correlated Gaussian perturbations. Numerically, the perturbations are independent of the coil discretization by considering additive perturbations modeled by Gaussian Processes (GPs). Similar approaches to model coil perturbations for stellarators have been presented in \citep{lobsien2020, wechsung2021singlestage}. Compared to \citep{wechsung2021singlestage}, instead of a squared exponential kernel with added periodicity, we take the period kernel as the basis for our perturbations. For the ease of notation, we omit the dependency on $i$ as the $i$-th coil in this subsection as without loss of generality we consider the derivation for one coil.

Let $\bm{X}(\xb,t)= (x(\x,t),y(\x,t),z(\x,t))^{T}$ be a parametric representation of a coil filament with $t\in[0,2\upi)$. We consider the distribution of the fabrications errors to be smooth along the entirety of the coil. As such we model them by zero-mean Gaussian random variables that effect every point on the coil. Equivalently, the coil filament $\bm{X}(\xb,t)$ is perturbed by a zero-mean Gaussian process (GP) 
\begin{align*}
(g_x(t),g_y(t),g_z(t))^{T}=:\bm{g}(t)\sim\GP(\bm{0},\kappa_{\text{per}}(t, t')), \quad t,t' \in [0,2\upi),
\end{align*} 
with a $2\upi$-periodic kernel $\kappa_{\text{per}}$. We assume that $\bm{g}(t)$ is an isotropic GP, i.e., the kernel $\kappa_{\text{per}}$ is only a function of distance of $t,t'$, and that $g_x(t),g_y(t),g_z(t)$ are independent of each other. Then the perturbed coil filament $\bm{X}_P(\x,t)$ is again a GP with mean $\bm{X}(\x,t)$
\begin{align}
    \bm{X}_{P}(\xb,t) :=\bm{X}(\xb,t) + \bm{g}(t) \sim\GP(\bm{X}(\xb,t), \kappa_{\text{per}}(t,t')), \quad t,t' \in [0,2\upi).
    \label{eq:gp_perturbation}
\end{align}
To compute relevant properties of the perturbed coil $\bm{X}_{P}(\xb,t)$ in FOCUS, we need a Fourier representation $\x_P(t)$ of $\bm{X}_P(\x,t)$. 
As we already know the Fourier representation of $\bm{X}(\x,t)$, we only need to calculate the Fourier representation of $\bm{g}(t)$.
% In the remainder of this section, we describe our method for computing the perturbations as a distribution over the Fourier coefficients $\xb$. 
 The $k$-th cosine and sine Fourier coefficients of the $x$-coordinate of the GP $g_x(t)$, are denoted $\hat{x}_{ck},\hat{x}_{sk}$, and can be calculated with 
\begin{align*}
    \hat{x}_{ck} &= \frac{1}{\upi}\int_0^{2\upi}g_x(t)\cos(kt)dt, \qquad
    \hat{x}_{sk} = \frac{1}{\upi}\int_0^{2\upi}g_x(t)\sin(kt)dt,
\end{align*}
with analogous forms for the $y$-$,z$-coordinates using $g_y(t),g_z(t)$.  We collect these Fourier coefficients into the vector $\U$. As $\bm{g}(t)$ is a GP and integration is a linear operation, it follows that $\U$ is a normally distributed random vector, see \citep{parzen1999stochastic}. The mean of $\U$ is zero, i.e. $\E[\U] = 0$ and the covariance can be written in terms of the Fourier coefficients of the GP kernel $\kappa_{\text{per}}$, \citep{parzen1999stochastic}, i.e., 
\begin{align}
    \cov[\hat{x}_{ck},\hat{x}_{sj}] = \frac{1}{\upi^2}
    \int_0^{2\upi}\int_{0}^{2\upi} \kappa_{per}(s,t)\cos(kt)\sin(js)\,ds\,dt.
    \label{eq:random_fourier_cov}
\end{align}
From the assumption that the $g_x(t),g_y(t),g_z(t)$ are independent of each other, it follows that their Fourier coefficients are independent as well. Thus, the covariance matrix $\bm{C}$ of the perturbation Fourier coefficients $\U$ is block diagonal, where each diagonal block has entries from \eqref{eq:random_fourier_cov}. Due to \eqref{eq:gp_perturbation}, a randomly perturbed coil can then be described in FOCUS' Fourier representation by
$
\x_P = \x + \U$, 
where $\bm{\U}\sim \mathcal{N}(\bm{0},\bm{C})$.

\begin{figure}
  %----------
  \subfigure{\label{fig:gp_lengthscale_0.1}
    \includegraphics[width=0.3\textwidth,height=0.23\textheight]{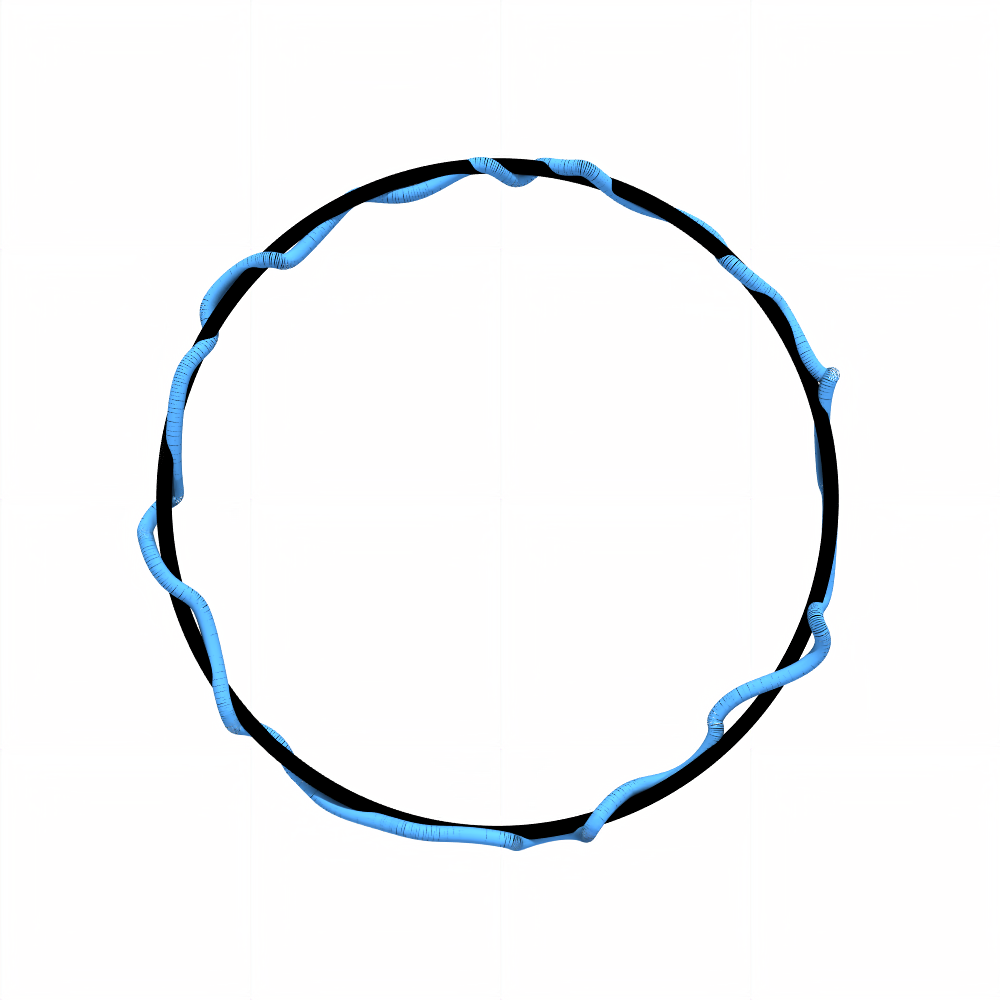}}
  %----------
  \subfigure{\label{fig:gp_lengthscale_0.5}
    \includegraphics[width=0.3\textwidth,height=0.23\textheight]{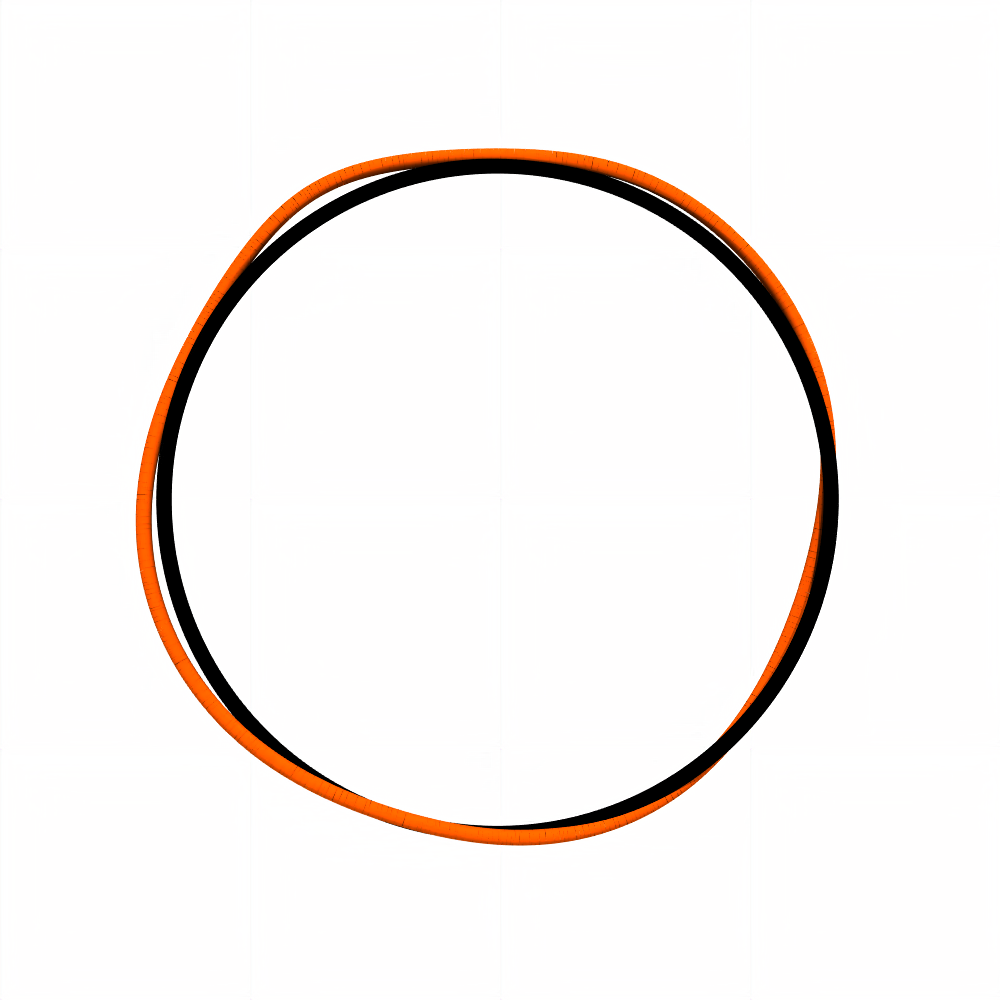}}
  %----------
  \subfigure{\label{fig:gp_lengthscale_1.0}
    \includegraphics[width=0.3\textwidth,height=0.23\textheight]{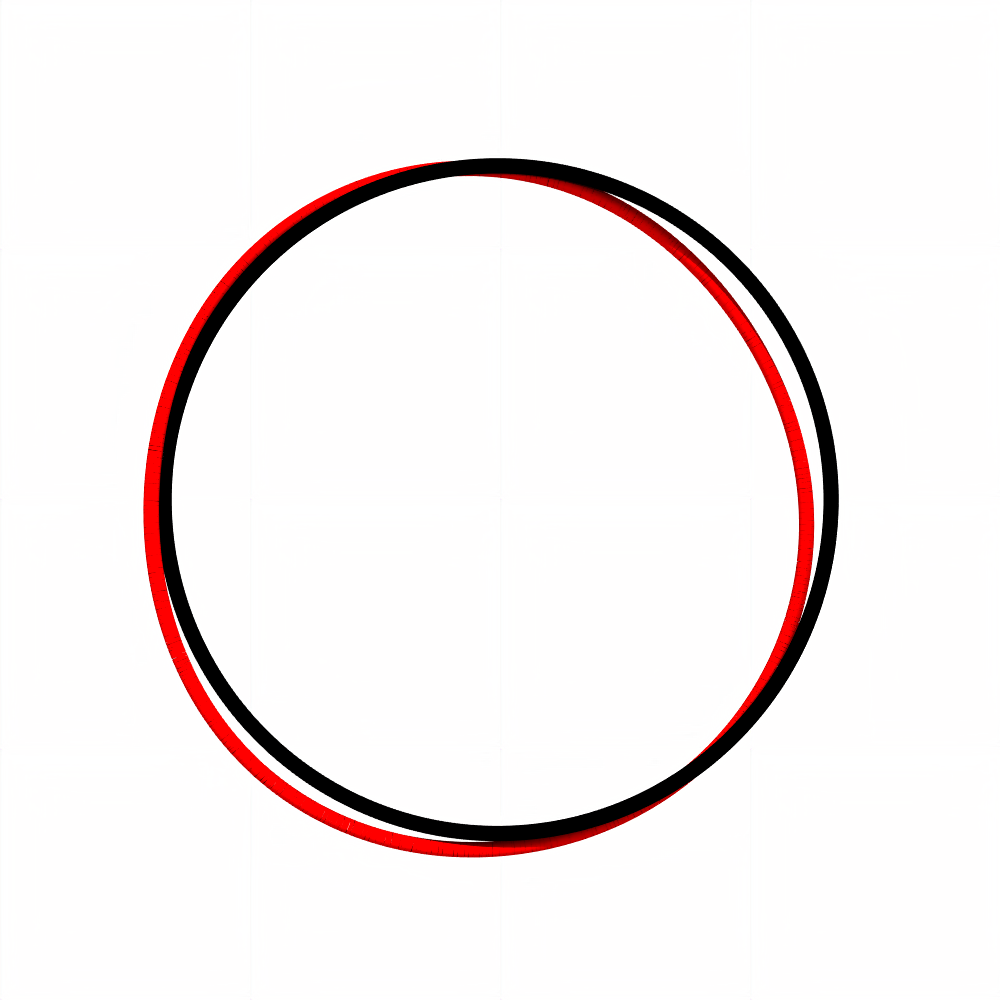}}
    \caption{Random perturbations of a circular coil generated from a GP with lengthscale $\ell=0.1$ (left), $\ell=0.5$ (center) and $\ell=1.0$ (right). Note that perturbations generated from a GP with a larger lengthscale have lower frequency oscillations.}
    \label{fig:gp_lengthscale}
\end{figure}

\subsubsection{The Kernel and Hyperparameters}
To ensure the GP perturbations are periodic, we use the $2\upi$-periodic kernel 
\begin{align*}
\kappa_{\text{per}}(t,t') = h\exp\left(-\dfrac{2\sin^2\left(\frac{|t-t'|}{2}\right)}{\ell^2}\right),  \quad t,t' \in [0,2\upi),
\end{align*}
where $h>0$ denotes the size of the perturbations and $\ell>0$ is the lengthscale of the GP. We set the hyperparameters $h, \ell$ such that at any point the mean norm squared of the multi-output GP equals the desired mean perturbation size $p$ squared, i.e.,  $\E[||\bm{g}(t)||_2^2] = p^2$. As $g_x(t),g_y(t),g_z(t)$ all have identical moments and a first moment of zero we simplify the equality to $3\var[g_x(t)] = p^2$ which yields $h = p^2/3$.

The lengthscale $\ell$ determines the frequency of the perturbations, as depicted in Figure \ref{fig:gp_lengthscale}. While we did not have any specific value given for $\ell$, we found $\ell=0.5$ to model realistic manufacturing errors such that we chose this value for our numerical experiments.

\subsection{Objectives}
\label{sec:objectives}
%In this section we describe a stochastic objective for encouraging robustness in the coil-induced magnetic field $f_B$ to fabrication errors in the coils. We also include a coil length regularization term to discourage excessiely long coils.

% While there are many physics and engineering penalty functions implemented in FOCUS such as, e.g., coil-to-coil separation energy, curvature, coil-plasma separation, 
The primary goal of the second stage of stellarator optimization is to find a set of coils which reproduce a target magnetic field while satisfying engineering targets, such as reasonable coil length and separation between coils. To this end, we chose our objective function $f$ to penalize the normal component of the magnetic field produced by the coils $f_B$ \eqref{Eq:average_squared_FE} and regularize the coil length $f_L$ \eqref{eq:length_objective} to discourage excessively long coils. 
We find that adding a simple penalty $f_L$ to the objective, rather than explicitly constraining the length, is sufficient to regularize coil length. While additional coil regularization functions such as curvature are available in FOCUS, using a length regularization proved to be sufficient. 

Not only do we seek a set of coils which generates the target magnetic field $\bv{B}$ - we simultaneously want to hedge against errors in the coil fabrication. To this end, we formulate the following stochastic objective 
\begin{align}\label{eq:def_f}
    f(\xb) = \E[\omega_B f_B(\xb+\U)] + \omega_L f_L(\xb), \qquad \x \in \R^{N},
    %\label{eq:risk-neutral}
\end{align}
where $\omega_B, \omega_L \in \R$ are weights of the respective objective, $\U\sim\mathcal{N}(\bm{0},\bm{C})$ is a multivariate normal random variable described in Section \ref{sec:uncertainty} and $\E$ denotes the expectation over $\U$. Referring to \eqref{eq:main_1}, we set the stochastic and the regularizing part of the objective to
\begin{align*}
f_\text{stoc}(\x) := \omega_B f_B(\xb), \qquad f_\text{reg}(\x):= \omega_L f_L(\xb).
\end{align*}
% Note, that we only choose to perturb the magnetic field as our main goal is to replicate the target magnetic field $\bv{B}$ as close as possible.
Note that we only penalize perturbations in the magnetic field penalty and not perturbations in the length penalty as fabrication errors that minorly increase coil length are not necessarily problematic from an engineering perspective.

%The dominant criterion $f_B$, equation \eqref{eq:f_bnormal}, in the objective $f$ addresses deviations in the magnetic field produced by a coil set $\xb$ from a target magnetic field. The target magnetic field is designed, in a previous stage of optimization not discussed here, to encode desirable confinement properties \todo{reference needed}. A secondary concern is the engineering and financial concern that coils must not be excessively long \todo{reference needed}. 

\subsection{Coil-to-Coil Separation Constraint}
\label{sec:constraints}
Engineering requirements state that a reasonable configuration must have sufficient spacing between all pairs of coils, see e.g.,  \citep{beidler1990physics}. In order to ensure that this condition is satisfied we include it as a constraint in our optimization model. The implication is that the coil-to-coil distance for adjacent coils is now \emph{bounded from below}, but any larger distance is allowed and will not add unnecessary penalties to the optimization model. We find that without enforcing this constraint coils come too close during numerical experiments.  
We model the minimum distance between coils $i$ and $i+1$ by
\begin{align}\label{eq:def_c}
c_i(\x) =  \min_{s,t \in[0,2\upi]^2} \|\bm{X}^i(\xb,s) - \bm{X}^{i+1}(\xb,t) \|_2^2.
\end{align}
Thus, for our optimization model in \eqref{eq:main_1}, we enforce the constraint

\begin{align} \label{eq:coil_constraint}
   \min_{s,t \in[0,2\upi]^2} \|\bm{X}^i(\xb,s) - \bm{X}^{i+1}(\xb,t) \|_2^2 \geq \epsilon_c^2 \ \ \ i \in\{1,...,N_C -1\}.
\end{align}
To handle the constraints \eqref{eq:coil_constraint} we discretize our coils into $N_{\text{seg}}$ segments and compute the minimum across a total of $N_{\text{seg}}^2$ discrete points for any adjacent coil pairs. Note, that it is important to also include the width of the coil into $\epsilon_c$ as FOCUS models coils as infinitely thin filaments.

In order to continue to use derivative-based optimization techniques we calculate the minimum distance and the derivative between a pair of coils with a smooth approximation to the minimum function, the so-called \emph{$\alpha$-quasimax} function, see \citep{lange2014smoothmin} (also known as \emph{LogSumExp} function). The $\alpha$-quasimax function approximates the minimum $\min(x_1, \ldots, x_n), \ x_i \in \R, i=1, \ldots, n$, by 
\begin{align}\label{eq:alpha_quasimax}
\mathcal{Q}_{\alpha}(x_1, \ldots, x_n) := \frac{1}{-\alpha} \log  \left(  \sum_{i=1}^{n} \exp\left(-\alpha x_i\right) \right),
\end{align}
with $\alpha >0$. Therefore, in the limit $\mathcal{Q}_{\alpha}(x_1, \ldots, x_n) \rightarrow \min(x_1, \ldots, x_n)$ for $\alpha \rightarrow \infty$ and the following bound holds
\begin{align*}
 \min(x_1, \ldots, x_n) - \frac{\log(n)}{\alpha} \le \mathcal{Q}_{\alpha}(x_1, \ldots, x_n) <  \min(x_1, \ldots, x_n).
\end{align*}

%\subsection{Final Optimization Model} \label{sec:final_model}

%Thus with the updates part of \eqref{eq:main_1} we arrive at the \emph{final optimization model} considered in this paper: we seek $\x \in \R^{N}$ such that
Thus, we arrive at the \emph{final optimization model} considered in this paper: we seek $\x \in \R^{N}$ such that
\begin{subequations} \label{Eq:main_stochastic}
\begin{align}
&\min_{\x \in \R^{N}} \ \   \E[  \omega_B  f_B(\x+\U)] +   \omega_L f_L(\x) \label{Eq:objective_stochastic}  \\
&      \min_{s,t \in[0,2\upi]^2} \|\bm{X}^i(\xb,s) - \bm{X}^{i+1}(\xb,t) \|_2^2 \geq \epsilon_c^2, \quad i=\{1,...,N_{c}-1\}. \label{Eq:constraint_stochastic} 
\end{align}  
\end{subequations}
In the next section, we present our efficient global optimization algorithm for the stochastic optimization of \eqref{Eq:main_stochastic}.

%% file: sections/global_optimization.tex
\section{Two-stage Global Stochastic Optimization}\label{Sec:global_high_dim_optimization}
%-----------------------------------------------------------------------------------------------------------------------------------------------
In this section, we introduce our global-to-local stochastic optimization algorithm for solving \eqref{Eq:main_stochastic}. To this end, we modify two methods, the global \emph{trust-region Bayesian optimization TuRBO} \citep{Turbo2019} method and the \emph{Adam} \citep{kingma2014adam} algorithm. The Bayesian derivative-free optimization routine TuRBO is designed to efficiently globally optimize a nonlinear function in a high-dimensional bound constrained space. As FOCUS provides derivative information we use DTuRBO \citep{padidar2021scaling}, which is a further development of TuRBO incorporating derivative information. To prioritize efficient exploration DTuRBO does not resolve minima to high orders of accuracy. Therefore, we apply a stochastic local optimization starting from the points we get from the final stage of DTuRBO. Thus our approach can be summarized in two stages:
\begin{enumerate}
	\item \emph{perform} an approximate \emph{global} optimization with DTuRBO and select a set of promising solution points,
	\item \emph{resolve} these solution points \emph{locally} utilizing a stochastic optimizer. 
\end{enumerate}
In order to provide choices for prospective future users of this two-stage approach, we introduce two options for a local stochastic optimizer, the Sample Average Approximation (SAA) method and the Adam algorithm enhanced with a \emph {novel control variate (AdamCV)} for variance reduction. This two-stage approach makes efficient use of the computational budget as the local optimizer will resolve the minima much more efficiently than DTuRBO. We start by describing the global optimization.
 
%To this end we apply a local stochastic optimization routine Adam with a novel control variate for variance reduction to refine the minima.

\subsection{Stage 1: Efficient Global Exploration of Design Space}\label{Sec:GlobalOpt}

%\subsubsection{Trust-Region Bayesian Optimization (TuRBO)}
In this section, we detail the global optimization by first describing TuRBO and subsequently commenting on the modifications made to arrive at DTuRBO. 

The TuRBO algorithm is a derivative-free method for global optimization of a nonlinear function across a high-dimensional hypercube $\Omega=[\text{lb},\text{ub}]^N$ with $\text{lb}, \text{ub} \in \R^{N}$ being a vector of lower and upper bounds for the design space variables. The TuRBO algorithm starts by performing a Latin hypercube sampling on $\Omega$ receiving a sample $X_{\text{init}}=[\x_1, \ldots, \x_{n_{\textnormal{sample}}}]$ of size $n_{\textnormal{sample}}\in \N$ and then evaluates the to be optimized function $f$ for each $\x  \in X_{\text{init}}$. Subsequently, $M$ local Bayesian optimization (BO) runs are started at the $M$ best points from $X_{\text{init}}$ by building local GPs within $M$ distinct rectangular trust-regions. At each iteration Thompson sampling, see \citep{10.1093/biomet/25.3-4.285}, is performed within each trust-region to generate a set of candidate points. A batch of the most promising of these candidate points is evaluated by tshe function $f$ and these function values are used to update the GP surrogates. Note, that the batch of evaluations is chosen from the union of candidate points across the trust-regions. Therefore, the evaluations are distributed to each trust-region by the region's predicted success through Thompson Sampling.
The trust-regions are centered around the best point found in the evaluation history, and are expanded or contracted by a factor of $2$ depending on consecutive successes/failures of decreasing the function value. 
A local BO will terminate after the trust-region reaches a minimum size. This indicates that the local BO is no longer making improvement and is near an optima. Precisely these approximate minima are gathered and passed on to the second stage of our approach for further refinement with the local optimizer. Then, a new Latin hypercube sampling of $\Omega$ is performed and a local BO is again started at the best point. The algorithm terminates when the computational budget is reached.
\\

%The advantage of using Thompson Sampling is that the trust-region optimizations progress at different rates and the more promising optimizations receive a larger portion of the computational budget. 
 %after $s$ consecutive successes (decreases in function value), or $f$ consecutive failures (no-decrease in function value), respectively. 
% The rectangular trust-region has side length proportional to the Gaussian Processes' lengthscale for that dimension. Note that the volume of the trust-region remains constant under a change of lengthscales. 
% In contrast to multi-start routines, TuRBO will only give a local optimization run computational resources if it predicts satisfactory decrease within the trust-region. This ensures that it does not squander resources in regions that are not promising. 
%By leveraging this multi-arm bandit strategy, Thompson Sampling, TuRBO can ensure that it is performing a near optimal search in terms of regret. 

The success of TuRBO in high dimensions is found in its ability to leverage multiple local surrogates and simultaneously efficiently distribute the computational budget across the local BO runs. While the common BO can only be used up to approximately 20 dimensions, see e.g. ~\citep{frazier2018tutorial}, TuRBO is designed for significantly higher-dimensional problems. \emph{DTuRBO} further improves on TuRBO by equipping it with a scalable method of incorporating derivatives into the GP models, which yields several advantages. For example, gradients encode the local descent direction ensuring that the surrogate is locally accurate and a decreasing direction can be found more easily. 
%Furthermore, each gradient evaluation is effectively equivalent to $N+1$ function evaluations, which ultimately reduces the total number of calls to FOCUS needed in an optimization. 
Internally DTuRBO uses a stochastic variational Gaussian process \citep{hensman2015scalable,jankowiak2020parametric} to scalably incorporate the high-dimensional gradient data by approximating it with a low dimensional representation, see \citep{padidar2021scaling}. For practical use, DTuRBO 
%is identical to TuRBO in the sense that it 
takes in a pair of noisy function and gradient evaluations 
%$(f(\x + \U),\nabla f(\x + \U))$ 
rather than only noisy function evaluations of $f$. The approximate minima gathered from DTuRBO  are passed as starting points for the local stochastic optimizers to be resolved further. %In the next section we describe such local routines.

\subsection{Stage 2: Local Stochastic Optimization}\label{Sec:LocalOpt}
Approximate minima found in stage one by the DTuRBO algorithm can further be refined by local stochastic optimizers. 
%For multiple high-quality minima, you may start the local optimizers at all local minima; a single high quality minima simply starts one optimization run from the best minima found. 
The benefit of using local stochastic optimization techniques is that they will converge to the minima faster than global optimizers as their main focus is on exploitation rather than exploration. Additionally, some of those methods provide convergence guarantees. Common choices for first-order techniques are variants of the stochastic gradient method (SGD) \citep{spall2005IntroStochasticSearch} or the application of non-stochastic optimizers within the SAA method, which has also been used in \citep{lobsien2020,wechsung2021singlestage}. The SAA approximation determines a fixed-accuracy approximation to the true stochastic objective, where the error in the approximation decays with the square root of the number of samples.  However, this approximation can be optimized in a relatively small number of steps of an algorithm like BFGS.  In contrast, stochastic gradient descent methods, including our AdamCV approach, converge to the optimum of the true stochastic objective using a large number of relatively inexpensive steps.  Which method is most appropriate depends strongly on how accurately the stochastic optimization problem should be solved.

In the following, we give an introduction to the Adam algorithm used in this work for the stochastic local optimization. Subsequently, we enhance Adam with a \emph{novel control variate}, which is a variance reduction technique to improve convergence. Additionally, we give a brief overview over the SAA method. 

%In the following, to increase readability and to focus on the main point - the global stochastic optimization - we restrict ourselves w.l.o.g. to the stochastic part of the optimization function $f:=f_{\text{stoc}}$ in this section. 

\subsubsection{Adam Algorithm}

In order to efficiently arrive at a well-refined solution, we use a variant of the \emph{Adam} algorithm \citep{kingma2014adam}. Adam is a popular stochastic optimization method in machine learning due to its improved performance over traditional stochastic gradient methods. Adam's success is largely due to the inclusion of a raw second moment estimator $\bm{v}_k$ of the stochastic gradient.
At each iteration Adam evaluates a small number $N_A$ (also called \emph{batchsize}) of gradients and averages to form a gradient estimator 
\begin{align}
\bm{g}^k &= \dfrac{1}{N_A}\sum_{i=1}^{N_A} \nabla f(\x_k+{\su}_{k,i}), \qquad k=1, \ldots, N_{\text{max}},
\label{eq:adam_grad}
\end{align}
where $N_{\text{max}}$ is the maximum number of iterations and $\su_{k,i}$ are $N_A$ realizations of the random variable $\U$. To stabilize gradient estimates, Adam employs exponential moving averages of the first and second raw moments of the stochastic gradient $\bm{m}_k, \bm{v}_k\in \R^{N}$ in each step
\begin{align*}
\bm{m}_k &= \beta_1 \bm{m}_k + (1-\beta_1)\bm{g}_k,  \qquad k=1, \ldots, N_{\text{max}},
\\
\bm{v}_k &= \beta_2 \bm{v}_k + (1-\beta_2) \bm{g}_k^2,  \qquad \ \,  k=1, \ldots, N_{\text{max}},
\end{align*}
where the parameters $\beta_1, \beta_2 \in [0,1]$ are usually set close to $1$. The first and second moments estimators $\bm{m}_k, \bm{v}_k$ are biased as they are initialized as the vector of all zeros. To correct the bias, the estimators $\bm{m}_k,\bm{v}_k$ are divided by $(1-\beta_1^k)$ and $(1-\beta_2^k)$ to create the bias corrected estimators $\hat{\bm{m}}_k,\hat{\bm{v}}_k \in \R^{N}$ 
\begin{align*}
\hat{\bm{m}}_k = \frac{\bm{m}_k}{(1-\beta_1^k)}  \qquad
\hat{\bm{v}}_k = \frac{\bm{v}_k}{(1-\beta_2^k)}  \qquad \ \,  k=1, \ldots, N_{\text{max}}.
\end{align*}
Utilizing these estimators Adam converges to a local minima of \eqref{Eq:main_stochastic} with the following step sequence
\begin{align*}
\x_{k+1} = \x_k - \eta_k \hat{\bm{m}}_k/(\sqrt{\hat{\bm{v}}_k}+\epsilon_A),
%\label{eq:adam_update}
\end{align*}
where $\epsilon_A \in \R^{+}$ is a small parameter to improve conditioning. To improve the convergence rate we use the decreasing step size sequence $\eta_k = \eta/(1+\sqrt{k}\gamma)$ where $\eta,\gamma$ are tunable parameters.
% (see appendix \ref{App:optimization_params}). 
 It is essential that the learning rates for Adam are well-tuned for good convergence.

\subsubsection{Adam with Control Variates (AdamCV)} \label{sec:adam_cv}

When the variance of the perturbations $\U$ become large it is beneficial to use \emph{variance reduction techniques} to reduce the variance of the gradient estimate $\bm{g}_k$. We follow \citep{wang2013sgdcontrolvariates} in developing such a variance estimator for $\bm{g}_k$ with control variates. By using a Taylor expansion of the FOCUS objective function we can derive an approximate stochastic gradient to $f$ with moments that are easy to calculate. This approximate gradient can be combined with the true gradient to create an unbiased estimator for $\bm{g}_k$ with a reduced variance. The first-order Taylor expansion of $\nabla f$ is $\nabla f(\bm{z}) \approx \nabla f(\x) + \bm{H}(\bm{z}-\x)$
where $\bm{H}$ denotes the Hessian matrix of $f$. 
% \begin{align}
% \label{Eq:Taylor_exp}
% f(\bm{z}) \approx f(\x) + \nabla f(\x)^T(\bm{z}-\x) + 0.5(\bm{z}-\x)^T\bm{H}(\bm{z}-\x),
% \end{align}
% Taking the gradient of \eqref{Eq:Taylor_exp} yields
% \begin{align*}
% \nabla f(\bm{z}) \approx \nabla f(\x) + \bm{H}(\bm{z}-\x),
% \end{align*}
Evaluating the Taylor expansion at $\bm{z} = \x+\U$ yields the control variate $\tilde{{g}}(\x)$
\begin{align}
\label{Eq:control_variate}
\tilde{{g}}(\x)= \nabla f(\x) + \bm{H}\U,
\end{align}
which is an estimator for $\E[\tilde{{g}}(\x)] = \nabla f(\x)$.
Combining the estimator $\tilde{{g}}$ \eqref{Eq:control_variate} with our original gradient estimator $\bm{g}_k$ \eqref{eq:adam_grad} leads us to a unbiased gradient estimator with lower variance
\begin{equation}
g(\x) = \nabla f(\x+\U) + \bm{A}(\tilde{{g}}(\x) - \E[\tilde{{g}}(\x)]),
\label{eq:controlvariate_estimator}
\end{equation}
where the choice of the diagonal matrix $\bm{A}\in \R^{N \times N}$ is detailed in the following.
This unbiased gradient estimator with lower variance can be applied to \eqref{eq:adam_grad} as a replacement for $\nabla f(\x+\U)$ to accelerate the convergence of the Adam routine. By following this approach, we ensure that the control variate is highly correlated with $\nabla f(\x+\U)$, such that we obtain the variance reduction with respect to the gradient estimates. 

As shown in \citep{wang2013sgdcontrolvariates} the optimal diagonal matrix $\bm{A}\in\R^{N\times N}$ is chosen in order to minimize the trace of the variance of $g(\x)$, i.e.,  
\begin{align*}
\bm{A} %&= \text{diag}(\cov(\tilde{\bm{g}}(\x),\nabla f(\x+\U)))/\text{diag}(\cov[\tilde{\bm{g}}(\x)),\tilde{\bm{g}}(\x)])
%\\
= \frac{\text{diag}\left(\cov(\bm{H}\U,\nabla f(\x+\U)) + \cov(\nabla f(\x+\U),\bm{H}\U)\right)}{2 \ \text{diag}\left(\bm{H}\U \U^{T}\bm{H}^T\right)},
\end{align*}
where we use \emph{diag} to indicate the diagonal entries of a matrix. 
%The numerator is approximated online with sample averages from the mini-batches. 
By using the optimal $\bm{A}$ in \eqref{eq:controlvariate_estimator} we arrive at a reduced variance for $\tilde{{g}}(\x_k)$ given by
\begin{align*}
\var[{\tilde{\bm{g}}(\x_k)}] = (1-\bm{\rho}^2)\var[{g}_k],
\end{align*}
where $\bm{\rho}\in\R^N$ has entries $\bm{\rho}_i = \corr(\partial_{\x_i} f(\x+\U),\tilde{\bm{g}}(\x_k)_i ),\ i = 1, \ldots, N$. 
As the analytic Hessian $\bm{H}$ of $f$ is not available in FOCUS we use an approximate Hessian $\bm{H}_k$
%\footnote{In particular, we use the approximate Hessian from the BFGS algorithm.} 
in optimization step $k$ according to the BFGS Hessian approximation \citep{nocedal2006numerical}
\begin{align*}
\bm{y}_{k-1} &:= \nabla f(\x_{k}) - \nabla f(\x_{k-1}),
\\
\bm{s}_{k-1} &:= \x_{k}-\x_{k-1},
\\
\bm{H}_{k} &= \bm{H}_{k-1} + \frac{\bm{y}_{k-1}\bm{y}_{k-1}^T}{\bm{y}_{k-1}^T\bm{H}_{k-1}\bm{s}_{k-1}}\
- \frac{\bm{H}_{k-1}\bm{s}_{k-1}(\bm{H}_{k-1}\bm{s}_{k-1})^T}{\bm{s}_{k-1}^T\bm{H}_{k-1}\bm{s}_{k-1}},
\end{align*}
with $\bm{H}_0$ initialized as the $N \times N$ identity matrix.
We find that equipping Adam with the control variate approach leads to rapid convergence rates in practice, particularly when warm-starting our optimization from a solution from stage one. We call the combination of the Adam algorithm enhanced with the control variate \emph{AdamCV}.%See Appendix \ref{App:optimization_params} for run parameters.

\subsubsection{Sample Average Approximation} \label{sec:saa_method}
The Sample Average Approximation (SAA) method, see \citep{shapiro2001monte,kim2015guide, kleywegt2002sample}, is a method of forming a non-stochastic approximation to the stochastic problem \eqref{Eq:main_stochastic} by using the Monte-Carlo method. For instance, in order to approximate the stochastic component  
$$\E[f_{\text{stoc}}(\x + \U)],$$
of the objective \eqref{Eq:main_stochastic}
, we pick $\{\bm{u}_i\}_{i=1}^{N_{\text{SAA}}}$, which are $N_{\text{SAA}}\in \N$ independent realizations of the random variable $\U$. These realizations $\{\bm{u}_i\}_{i=1}^{N_{\text{SAA}}}$ are taken to form the approximation 
\begin{align*}
f_{\text{SAA}}(\x) = \frac{1}{N_{\text{SAA}}}\sum_{i=1}^{N_{\text{SAA}}} f_{\text{stoc}}(\x + \bm{u}_i),
\end{align*}
with analogous form for the gradient approximation. This objective is straightforward to implement and can be minimized efficiently with standard non-stochastic optimizers such as BFGS \citep{nocedal2006numerical} since the samples $\bm{u}_i$ are kept fixed. Solving the SAA comes with large sample size guarantees. In the limit as the sample size $N_{\text{SAA}} $ approaches infinity, the order of convergence is $\mathcal{O}(1/\sqrt{N_{\text{SAA}}})$ as for standard Monte-Carlo methods.
% , the minima of the sample averaged problem will approach the minima of the stochastic problem \todo{need citation}. In contrast to stochastic optimizers such as Adam, 
For a finite batchsize $N_{\text{SAA}}$, the minima of the SAA may not converge to minima of the stochastic problem. So it is recommended that any minima to the SAA is re-evaluated under the stochastic objective to estimate the ``out-of-sample'' performance.

%% file: sections/experiments.tex
\section{Stellarator Experiments}\label{Sec:Experiments}

In this section, we present numerical results for a W7-X configuration. In the following, we detail our model data and perform the global-to-local stochastic optimization described in Section \ref{Sec:global_high_dim_optimization} to find multiple stochastic minima. To better differentiate between the global and the local optimizer, we perform a comparison of the local optimizers SAA and AdamCV initialized at equispaced circular coils in Section \ref{sec:exp_local}. Then, in Section \ref{sec:exp_global}, we use our two-step global optimization method to find multiple promising stochastic minima.  
 
\begin{center}
\centering
\begin{table}
\begin{tabular}{| p{2.3cm}| C{2.3cm}|C{2.3cm}|C{2.3cm}|C{2.3cm}|C{2.3cm} |}
 %\hline
% \multicolumn{7}{|c|}{Optimization parameters} \\
 \hline
 Parameter & $\omega_B$ & $\omega_L$ & $\epsilon_c$ & $L_{i}^{\text{target}}$ & $\ell$ \\
 \hline
Value   & $100.0$  & $0.5$ & $0.23$ & $8.0$ & $0.5$  \\
 \hline
\end{tabular}
\caption{Optimization parameter used in numerical experiments: weights in FOCUS objective function $\omega_B, \omega_L$, target length $L_{i}^{\text{target}}$, coil-to-coil separation $\epsilon_c$ and lengthscale $\ell$.}
\end{table} \label{Table:focus_opt_params}
\end{center}

\subsection{Model Data} \label{sec:model_data}
%\todo{Misha: The average absolute deviation from the central filament in the CAD design was less than $3$mm \citep{andreeva2015tracking}, satisfying the engineering tolerances. - do we really need this here?}
We perform experiments on a W7-X configuration, see e.g. \citep{klinger2019overview} for a description of W7-X. The stellarator consists of five distinct modular coils, such that after applying the stellarator symmetry as well as 5-field period symmetry we arrive at a total of 50 coils. For this study, we find setting the number of Fourier modes to $N_F=6$ to be sufficient, which results in a state dimension of $N = 195$. 
%In our numerical experiments, we run FOCUS with the objective function parameters given in table \ref{Table:focus_opt_params}. 
The weights used in our experiments as well as other optimization parameters for FOCUS can be found in Table \ref{Table:focus_opt_params}. Heuristically, we found that these weights struck a nice balance between minimizing the field error and finding smooth coils. Moreover, we used $N_{\text{seg}}=64$ segments per coil as well as $N_{\text{theta}}=N_{\text{zeta}} = 64$ nodes in either discretization direction of the plasma boundary. 
In order to enforce a reasonable coil-to-coil constraint for all pairs of adjacent coils, we have to take a minimum distance and the width of the coils into account. The latter is crucial as FOCUS models coils as infinitely thin filaments. Therefore, we motivate the value of $\epsilon_c$ by coil separation distances for the W7-X candidate configurations HS-5-7 and HS-5-8 given in \citep{beidler1990physics}. The minimum distance between the coils for the candidate configurations were $0.06$m and $0.04$m, respectively. The average lateral coil width of the coils was $0.18$m, such that we chose the coil-to-coil separation to be at least $\epsilon_c = 0.23 = \frac{1}{2}(0.04+0.06) +0.18$m in all of our experiments. The coil-to-coil separation constraints were included in the model via a quadratic penalty method, see \citep{nocedal2006numerical}. Thus all optimizations were performed on the penalty objective with $\lambda \in \R^{+}$
\begin{align*}
f_\text{pen}(\x) := f(\x) + \lambda\sum_{i=1}^{N_C - 1} \min[c_i(\x) - \epsilon_c^2,0]^2,
\end{align*}
with $f(\x)$ given in \eqref{eq:def_f} and $c_i(\x)$ \eqref{eq:def_c}. A $\lambda$ value of $100$ was found to be sufficient in consistently achieving constraint satisfaction. Moreover, we set $\alpha=10 000$ for the $\alpha$-quasimax function in \eqref{eq:alpha_quasimax}.

%\todo{Outline: to do}
%\begin{enumerate}
%	\item section: 5.2 local comparison: to choose optimizer - run AdamCV + SAA - detail how optimizer handle stochasticity and thus measure efficiency - which one to choose for problem at hand. detail pros and cons 
%	\begin{itemize}
%		\item Adam more difficult to tune if no prior experience
%		\item more is expensive - usually don't know how to set it
%		\item con: might need high batch size a priori SAA what
%	\end{itemize}
%\end{enumerate}

\subsection{Numerical Results for Local Optimization}\label{sec:exp_local}
We compare the AdamCV algorithm described in Section \ref{sec:adam_cv} to the BFGS algorithm applied to SAA \ref{sec:saa_method}. 
%For this comparison, we chose the perturbation size to be $p=10$mm, but we ran this experiment for different perturbation sizes $p =2$mm, $p= 5$mm, $p=20$mm and made similar observations. 
We initialized the local optimization of \eqref{Eq:main_stochastic} from a configuration of equispaced circular coils using the perturbation size $p=10$mm for all experiments in this section. We ran this experiment for the different perturbation sizes $p =2$mm, $p= 5$mm, $p=20$mm and made similar observations.

%Version 1:\\

\begin{figure}
	%----------
	\subfigure[\label{fig:coils_inboard_SAA_Adam} SAA $p=10$mm.]{
		\includegraphics[width=0.50\textwidth,height=0.23\textheight]{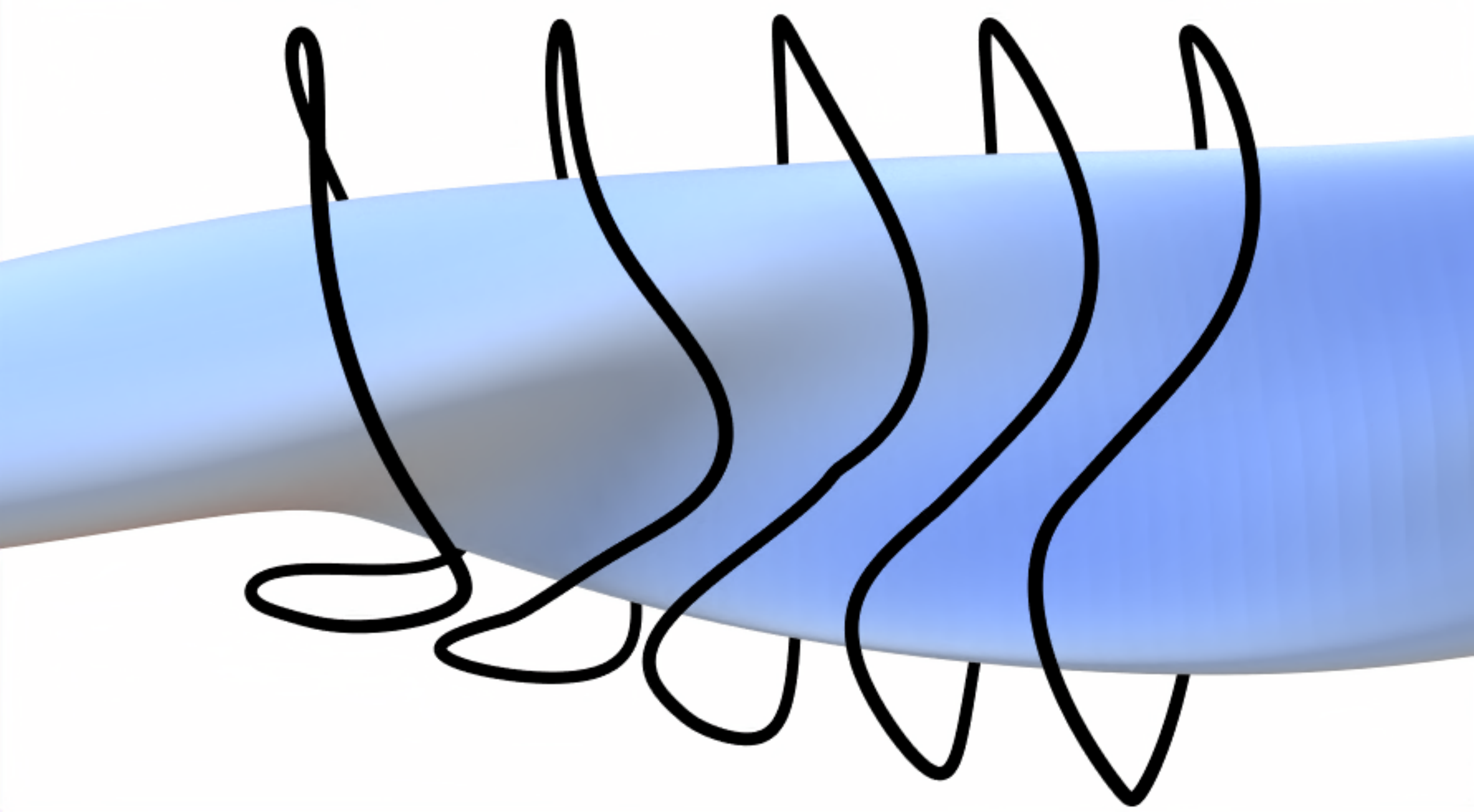}}
	%----------
	\subfigure[\label{fig:coils_Outboard_SAA_Adam} AdamCV $p=10$mm.]{
		\includegraphics[width=0.50\textwidth,height=0.23\textheight]{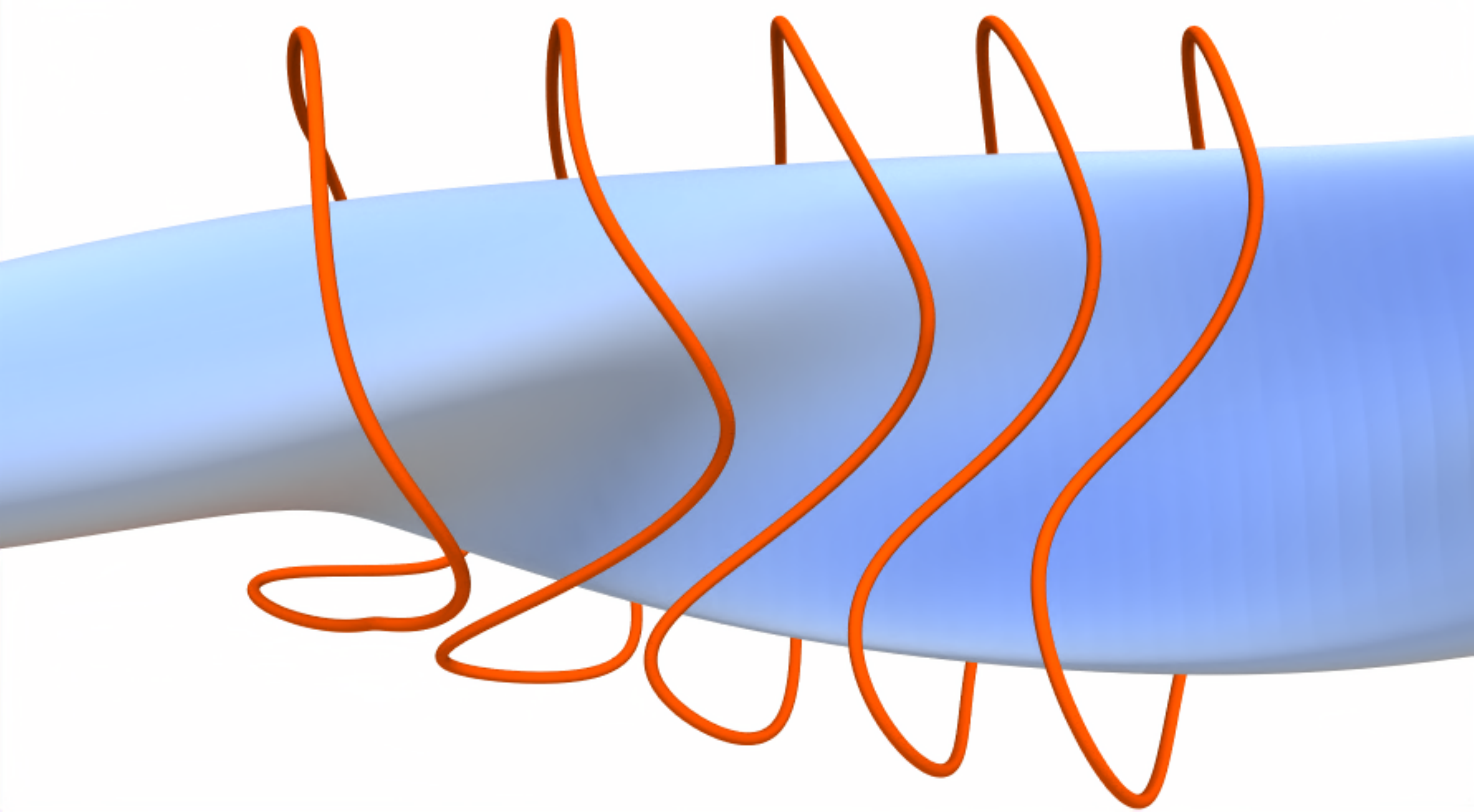}} \\
	%----------
%	\centering
%	\includegraphics[scale=0.3]{figures/w7x_jf_adamcv_opt_p_001_seed_20210824223605.pdf}
	  \caption{Final coil configuration of local stochastic optimization: optimized with SAA (left) and with AdamCV (right). } 
\label{Fig:W7-X_coils_local} 
\end{figure}

In order to compare the two algorithms as fair as possible, we set the number of \emph{gradient evaluations per step} to 10 and the \emph{maximum number of gradient evaluations} to 50 000. The remaining parameters of AdamCV are set to $\eta=0.04$, $\gamma=0.1$, $\beta_1=\beta_2=0.95$, $\epsilon_{A}=10^{-10}$. The SAA approximation was optimized with the deterministic SciPy optimizer BFGS \citep{virtanen2020scipy} and was restarted with new sample values $\bm{u}_i$ once a minimum for a fixed sample set had been reached. Due to the restarting, we find that a sample size of $10$ is indeed enough for the SAA algorithm, as we find very similar values for e.g., batchsize $100$.  We plot the final coil sets found by SAA and AdamCV in Figure \ref{Fig:W7-X_coils_local} and find the coils to be similar. We compute the mean squared curvature values
\begin{align*}
\frac{1}{N_C}\sum_{i=1}^{N_C} \int_{0}^{2\upi} \kappa_i^2(t) \, dt,
\end{align*} 
with $\kappa_i$ being the curvature for coil $i$ for the respective coil sets: the SAA coils obtain a squared curvature value of $5.313$, while the coils optimized with AdamCV are a bit smoother with a mean squared curvature value of $3.134$. 
Additionally, the BFGS needs to evaluate the objective function as many times as the gradient such that we have an additional 50 000 function evaluations adding to the cost.   
% In terms of computing time for our setting, the SAA algorithm takes about of $83$ minutes longer.

% In the following, we focus on the evaluation under uncertainty of the function value and magnetic field. 
We measure the quality of the coil sets by looking at 3 measures: the stochastic objective function $f(\x)$ \eqref{Eq:objective_stochastic}, the normal field error $f_B(\x)$\eqref{Eq:average_squared_FE} and the stochastic normal field error $\E[f_B(\x+\U)]$. We provide an overview of our findings in Table \ref{Table:FE}. We find that when optimizing with the AdamCV algorithm we arrive at a similar stochastic function value, with the AdamCV algorithm providing a $0.4\%$ smaller value than the SAA. When we evaluate the stochastic component of the objective function, i.e., the stochastic normal field error, AdamCV finds a $2.1\%$ smaller stochastic field error value than SAA. 

% Concluding the comparison for this setting in FOCUS, on the one hand 
We find that AdamCV arrives at slightly lower stochastic function values/field error, at a improved computational expense as it does not need to evaluate the function values additionally to the gradient evaluations. 
This success is in part due to proper selection of the learning rate parameters $\eta,\gamma$.
% Nevertheless, at the same time some experience with tuning the parameters of AdamCV is helpful to achieve good convergence. 
The SAA procedure performs similarly well, is easy to implement, and works well with only having to choose the sample size. We recommend restarting the SAA optimization with a new batch of samples after a run converges, as we see an improvement throughout the subsequent runs. 
In our experience, both methods can work well in the local refinement step of our two-stage approach. We choose to use the AdamCV algorithm due to it's improved performance over the SAA approach.

\begin{center}
\begin{table} \label{Table:FE}
\begin{tabular}{| p{3.5cm}|p{4.6cm}|p{2.3cm}|p{4.6cm} | }
 \hline
% \multicolumn{4}{|c|}{W7-X} \\
 %\hline
Coil Configuration & Stochastic Obj. Function & Field Error& Stochastic Field Error  \\
\hline
%Reference & $3.993\times 10^{-2}$ &$3.331\times 10^{-4}$  & -- \\
%Circular coils & $2.431\times 10^{+1}$ &$1.400\times 10^{-1}$ & &   \\
%TuRBO-Adam 2mm & $4.092\times 10^{-2}$ &$3.140\times 10^{-4}$  & --  \\
SAA 10mm  &$4.728\times 10^{-1} \pm[1.10 \times 10^{-2}] $  & $1.472\times 10^{-3}$& $1.972\times 10^{-3}\pm[1.10 \times 10^{-4}]$ \\ \hline
AdamCV 10mm &$4.707\times 10^{-1}\pm[1.05 \times 10^{-2}] $  & $1.424\times 10^{-3}$& $1.930\times 10^{-3}\pm[1.05 \times 10^{-4}]$\\
%TuRBO-Adam 20mm & $4.805\times 10^{-2}$ &$4.793\times 10^{-4}$  & -- \\
 \hline
\end{tabular}
\caption{Values of stochastic objective function \eqref{Eq:objective_stochastic}, normal field error \eqref{Eq:average_squared_FE} and stochastic normal field error $\E[f_B(\x+\U)]$ for W7-X. The stochastic values are computed using perturbation size $p=10$mm and we are averaging over 1000 realizations of $\U$. For the stochastic values we include the 95$\%$ confidence interval after the function value.}
\end{table}
\end{center}

\subsection{Numerical Results for Global Optimization}\label{sec:exp_global}
%\todo{Earlier in section: describe how we chose the perturbation sizes and how 2mm is too small. Say we chose a w7-x configuration, describe the use of a penalty formulation.}
In this section, we describe the global stochastic optimization of a W7-X configuration with model data given in Section \ref{sec:model_data}. 
%Here we describe the global optimization of the coil configurations for a W7-X like configuration through the model described in \todo{ref model section and model data}. 
For our two-stage approach detailed in Section \ref{Sec:global_high_dim_optimization}, we use the pair DTuRBO and AdamCV for the efficient global search and local refinement steps, respectively. %We chose AdamCV over SAA because AdamCV guarantees convergence to a local optima of \todo{reference main} with any batch size and converged relatively efficiently in our initial experiments \todo{reference previous section}. 
In our experiments, we set the average perturbations amplitudes to $p = 5$mm and $p = 10$mm. The global exploration algorithm DTuRBO was given a maximum number of 100 000 evaluations, a batchsize of $100$ and $200$ initial evaluations. 

The bounding boxes for DTuRBO should be set large enough such that there is enough flexibility in the design space, while not so large as to capture poor regions of the design space. 
% At the same time, we are using Fourier series to represent the coils such that the box constraints should get narrower for higher-order Fourier modes. 
As the design variables are Fourier coefficients the box constraints should get narrower for higher-order Fourier modes.
To this end, the lower and upper bounds were computed using the variance of the perturbations as an approximate lengthscale.
We set the bounding boxes for DTuRBO to be centered around $\x_0$, which denotes circular coils of radius $1.5$m, resulting in $\text{lb},\text{ub} = \x_0 \pm\delta\var[\U]$. The scalar $\delta = \frac{1.5}{2\var[U_0]}$ resizes the box width such that the translational modes have perturbations bounded by $1.5/2$ m, where $\var[U_0]$ is the perturbation variance to the translational mode. %The values of $\alpha$ results in $\alpha \approx$ for the $p=5$mm perturbation and in $\alpha \approx$ for the $p=10$mm perturbation. %are the descicion variables for the circular coils and $\alpha = $\todo{need $\alpha$ value} is a scaling constant chosen large enough to observe large penalty values \todo{finish thought}. 

\begin{figure}
	%----------
	\subfigure[\label{fig:coils_inboard_SAA_Adam} DTuRBO-AdamCV $5$mm.]{
		\includegraphics[width=0.50\textwidth,height=0.28\textheight]{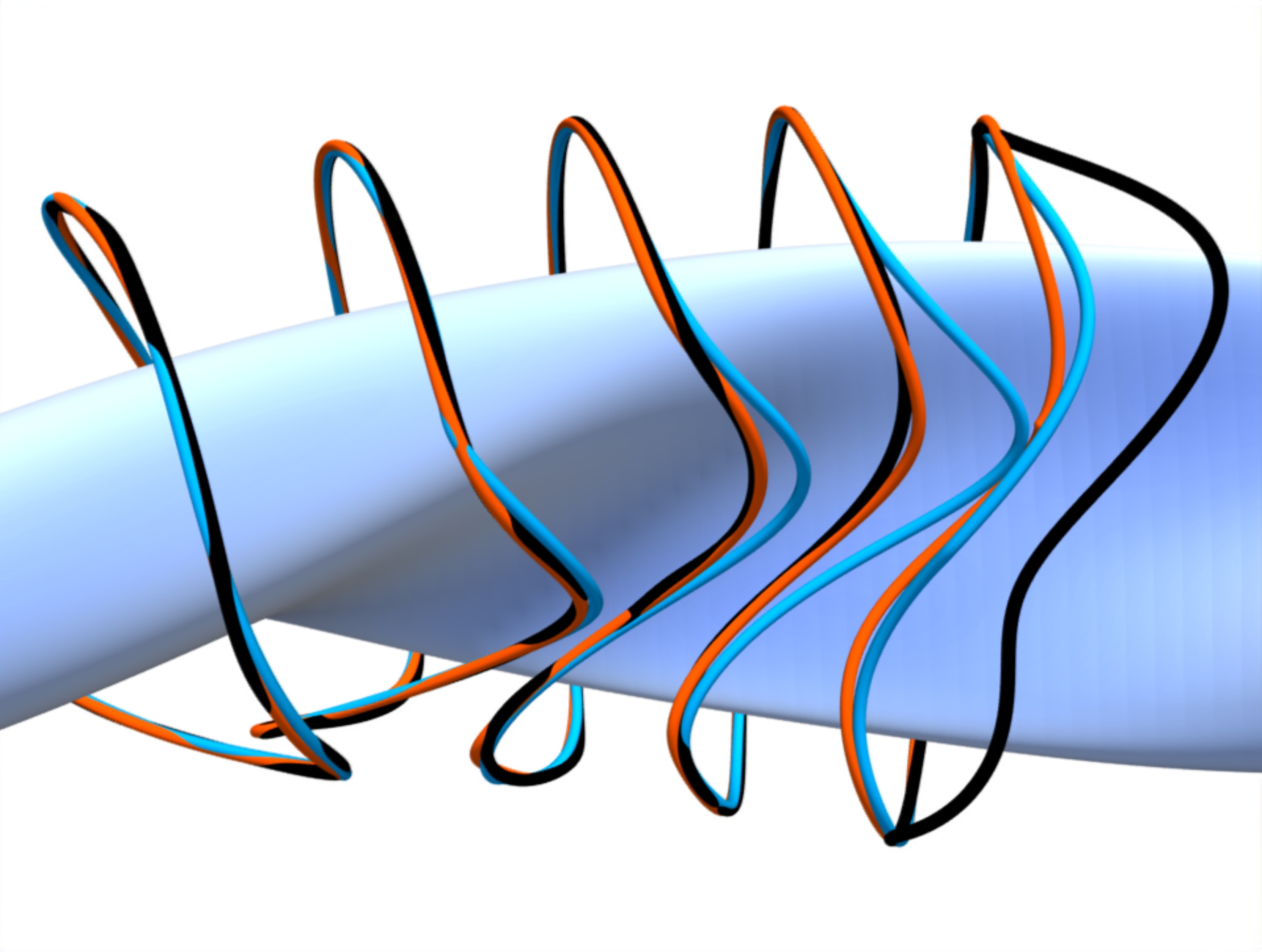}}
	%----------
	\subfigure[\label{fig:coils_Outboard_SAA_Adam} DTuRBO-AdamCV $10$mm.]{
		\includegraphics[width=0.50\textwidth,height=0.28\textheight]{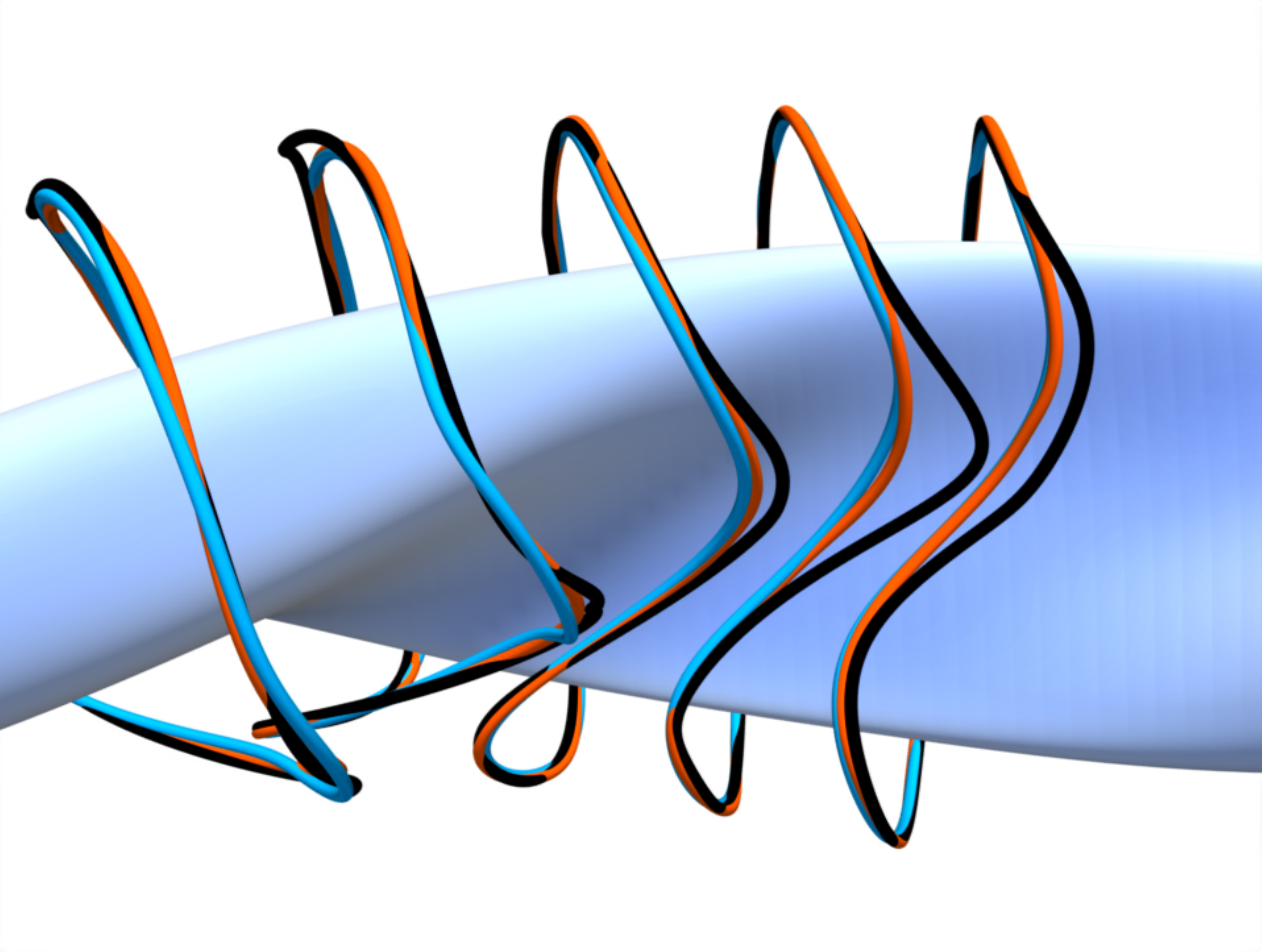}} \\
	\caption{Final coil sets for the optimization with DTuRBO and AdamcV. Left: $p=5$mm; right: $p=10$mm. Corresponding field error can be found in Table \ref{Table:Global}. For each plot, we sorted and colored the coils according to their stochastic objective value with respect to Table \ref{Table:Global}: low (red), medium (blue), high (black). In plot (a), the two rightmost blue coils appear to be very close. This is deceptive as the rightmost coil extends outward while the other passes behind it. }
\label{Fig:W7-X_coils_final_minima}
\end{figure}

Using this optimization setting, DTuRBO finds around 15 approximate stochastic minima, which were subsequently resolved with the minimizer AdamCV with a maximum number of $2000$ iterations with a batchsize of $10$ and parameters  $\eta=0.001$, $\gamma=0.01$, $\beta_1=\beta_2=0.95$, $\epsilon_{A}=10^{-10}$. Using this optimization setting, on average, the combined optimization routine DTuRBO \& AdamCV found 8 approximate minima within 116 000 evaluations, which had low enough stochastic objective value/field error to use them for further study of physical properties. 
In Figure \ref{Fig:W7-X_coils_final_minima}, we show three final coil sets found by the global optimization for the different perturbation sizes. Although it seems like the coils are close in the plots, the minimum distance is satisfied in all configurations. All of these coil sets achieved a low normal field error $f_B(\x)$ and stochastic normal field error $\E[f_B(\x +\U)]$, as seen in Table \ref{Table:Global}. We also compute the 95\% confidence interval of each coil configuration and find the size to be at least a magnitude smaller than the respective value, varying slightly with respect to the coil configuration. %We found that the coils found under a $s=2$mm average perturbation amplitude were similar to those optimized by non-stochastic optimization.

Comparing to the available literature we find that our total costs for arriving at an approximate stochastic global minimum is less than $0.1 \%$ compared to the evaluation budget in \citep{lobsien2020}, where solely local stochastic optimization has been used. Here, we assume that function evaluations in the respective codes take a similar amount of time/resources and we use a 5-to-1 conversion factor to convert the time for gradient evaluations to the time for function evaluations.\footnote{We run our experiments on an Ubuntu 20.04 cluster with MPI using 14 cores. Across $10^4$ calls to FOCUS, the mean time for the function evaluations is around 0.1s and for the gradient evaluations is around 0.5s.} We attribute this improvement in efficiency to a judicious choice of algorithms and the availability of gradients in FOCUS. %The runtime could also be effected by the nonlinearity and effects of stochasticity in the objective. This goes to show that Stellarator coil configurations can be \textit{globally} optimized relatively efficiently.

% In order to relate the cost of function values and gradient evaluations of \ref{Eq:main_stochastic} to each other we ran 10 000 evaluations of each and computed the mean time. We found that multiplying the gradient evaluations with a factor of 5 is accurate for our numerical experiments

\begin{center}
\begin{table} \label{Table:Global}
\begin{tabular}{| p{3.5cm}|p{4.6cm}|p{2.3cm}|p{4.6cm} | }
 \hline
% \multicolumn{4}{|c|}{W7-X} \\
 %\hline
Coil Configuration & Stochastic Obj. Function & Field Error& Stochastic Field Error  \\
\hline
D-ACV-5 (red) &$4.360\times 10^{-1} \pm[2.86 \times 10^{-3}] $  & $1.472\times 10^{-3}$& $1.599\times 10^{-3}\pm[2.86 \times 10^{-5}]$ \\ \hline
D-ACV-5 (blue)   &$4.795\times 10^{-1} \pm[2.78 \times 10^{-3}] $  & $1.635\times 10^{-3}$& $1.758\times 10^{-3}\pm[2.78 \times 10^{-5}]$ \\ \hline
D-ACV-5 (black)  &$4.811\times 10^{-1} \pm[3.19 \times 10^{-3}] $  & $1.584\times 10^{-3}$& $1.723\times 10^{-3}\pm[3.19 \times 10^{-5}]$ \\ \hline
D-ACV-10 (red) &$4.701\times 10^{-1} \pm[1.06 \times 10^{-2}] $  & $1.419\times 10^{-3}$& $1.925\times 10^{-3}\pm[1.06 \times 10^{-4}]$ \\ \hline
D-ACV-10 (blue) &$5.102\times 10^{-1} \pm[1.17 \times 10^{-2}] $  & $1.638\times 10^{-3}$& $2.155\times 10^{-3}\pm[1.17 \times 10^{-4}]$ \\ \hline
D-ACV-10 (black)  &$5.157\times 10^{-1} \pm[9.78 \times 10^{-3}] $  & $1.549\times 10^{-3}$& $2.082\times 10^{-3}\pm[9.78 \times 10^{-5}]$ \\ \hline
\end{tabular}
\caption{Values of stochastic objective function \eqref{Eq:objective_stochastic}, normal field error \eqref{Eq:average_squared_FE} and stochastic normal field error $\E[f_B(\x+\U)]$ for 6 different coil configurations for W7-X shown in Figure \ref{Fig:W7-X_coils_final_minima}. The stochastic values are computed using the respective perturbation size, and averaged over 1000 realizations of $\U$. For the stochastic values we include the 95$\%$ confidence interval after the function value. The color adjacent to the coil configuration denotes the corresponding colored coil set in Figure \ref{Fig:W7-X_coils_final_minima}. All coils have been optimized with DTuRBO and AdamCV (D-ACV) and the number in the coil configuration indicates the perturbation size.}
%\caption{Values of stochastic objective function $f$ \eqref{Eq:objective_stochastic}, field error $f_B$ \eqref{Eq:average_squared_FE} and stochastic field error $\E[f_B(\x+\U)]$ for multiple globally optimized solutions found under varying perturbation amplitudes $s$. The stochastic values are computed using the respective perturbation size of the experiment and we are averaging over \todo{xxx} realizations of $\U$. \todo{fill out table with new values}}
\end{table}
\end{center}

%% file: sections/conclusion.tex
%!TEX root = ../main.tex

%------------------------------------SECTION: CONCLUSION --------------------------------------------------------
\section{Conclusion and Future Work}\label{Sec:Conclusion}
%--------------------------------------------------------------------------------------------------------------------------------

% We introduced a novel efficient stochastic optimization algorithm for coil optimization in FOCUS. In particular, our algorithm is divided into two steps: a global exploration phase, where we build a surrogate model to ensure efficiency. In a second step, we resolve the minima found in step one with the AdamCV optimizer, which is the Adam optimizer enhanced with a control variate approach to perform variance reduction and speed up computations. We performed experiments on a W7-X configuration and find that compared to available Monte-Carlo-based methods in the literature, our algorithm is finding risk minima with comparable average and maximum field errors while at the same time reducing the computational cost to around $0.1\%$. \\

In this paper we develop a stochastic optimization model for stellarator coil configurations in order to hedge against fabrication errors in the construction.
Our model considers the effects of normally distributed coil fabrication uncertainties on the normal field error, a length regularization and a coil-to-coil separation distance constraint. Our novel global-to-local approach leverages the efficient high-dimensional derivative-based Bayesian optimizer DTuRBO, as well as SAA and the Adam algorithm equipped with control variates to perform an efficient global exploration. In our numerical experiments for a W7-X-like configuration, we found many satisfactory minima at a low computational expense, approximately less than $0.1\%$ of previous work. Note, that previous work only addresses local stochastic optimization, whereas we perform global stochastic optimization.

Possible further directions of work include the investigation of other objective functions which show high sensitivity to coil errors. For instance, \citep{andreeva2015tracking,kremer2007creation} showed the magnetic island width to be a quantity highly affected by errors in the coils. Another direction might be to perform this global-to-local approach with e.g., different physical properties using different codes such as SIMSOPT \citep{landreman2021simsopt} or PyPlasmaOpt \citep{wechsung2021singlestage}. Both of these codes include derivative information, which has been indispensable in this work. Furthermore, the model posed here only considers fabrication uncertainties whereas coil placement and alignment uncertainties, considered in \citep{kremer2007creation}, are yet another source of error with a distinct distribution. Future work could investigate the distributional assumptions of the models. \\

% Next, the Gaussian model for the uncertainty distribution is has only modest justification. Measurements of the errors were taken in \todo{cite andreeva} and could be used to generate a distribution or guide a data-driven approach to the stochastic optimization. In addition our model could benefit from experimentation with other objectives measuring stochastic robustness, such as a robust optimization or probabilistically robust approach \todo{citation}. These approaches would come with further gaurantees of how the uncertainty would effect the selected solutions.

% Lastly, the model is dependent on a series of parameters such as the size of the perturbations $m_s,m_p$, the GP lengthscale $l$, mesh sizes $n_s,n_p$, constraint parameters $\epsilon_c,\epsilon_p,L_0,\delta$. A sensitivity analysis should be performed to understand the dependence of the optimal solutions on these parameters. For constraint parameters that can be performed through the KKT conditions \citep{nocedal2006numerical}, while for perturbation distribution parameters we can look to the Fischer Information, and discrete mesh parameters we can manually investigate the fidelity. In addition to a sensitivity analysis, the solution sets generated by the model could use further investigation through a physics analysis similar to \citep{lobsien_drevlak_kruger_lazerson_zhu_pedersen_2020}. 

\noindent{\textbf{Acknowlegements}}

Silke Glas and David Bindel acknowledge support from the Simons Foundation in the Collaboration on Hidden Symmetries and Fusion Energy. In addition David Bindel has been supported by the NSF under (1934985) and Ariel Kellison has been supported by the DOE CSGF under (DE-SC0021110). All authors thank M. Landreman for fruitful discussions and support, J.-F. Lobsien for sharing his FOCUS W7-X input file, and C. Zhu for his help with the FOCUS code.